\documentclass[conference]{IEEEtran}
\IEEEoverridecommandlockouts

\usepackage{cite}
\usepackage[utf8]{inputenc}
\usepackage{amssymb}
\usepackage{graphicx}
\usepackage[caption=false]{subfig}

\newcommand{\cm}{\checkmark}

\begin{document}

\title{Testing Deep Learning Models: A First Comparative Study of Multiple Testing Techniques
}

\author{\IEEEauthorblockN{Mohit Kumar Ahuja}
\IEEEauthorblockA{\textit{Simula Research Laboratory} \\
Oslo, Norway \\
mohit@simula.no}  %
\and
\IEEEauthorblockN{Arnaud Gotlieb}
\IEEEauthorblockA{\textit{Simula Research Laboratory} \\
Oslo, Norway \\
arnaud@simula.no}  %
\and
\IEEEauthorblockN{Helge Spieker}
\IEEEauthorblockA{\textit{Simula Research Laboratory} \\
Oslo, Norway \\
helge@simula.no}  %
}

\maketitle

\begin{abstract}
Deep Learning (DL) has revolutionized the capabilities of vision-based systems (VBS) in critical applications such as autonomous driving, robotic surgery, critical infrastructure surveillance, air and maritime traffic control, etc. By analyzing images, voice, videos, or any type of complex signals, DL has considerably increased the situation awareness of these systems. At the same time, while relying more and more on trained DL models, the reliability and robustness of VBS have been challenged and it has become crucial to test thoroughly these models to assess their capabilities and potential errors. To discover faults in DL models, existing software testing methods have been adapted and refined accordingly. In this article, we provide an overview of these software testing methods, namely differential, metamorphic, mutation, and combinatorial testing, as well as adversarial perturbation testing and review some challenges in their deployment for boosting perception systems used in VBS. We also provide a first experimental comparative study on a classical benchmark used in VBS and discuss its results.
\end{abstract}

\section{Introduction}
The performance of deep learning (DL) has been improving rapidly and the industrial deployment of DL models is increasing in various domains, such as computer vision, video analysis, natural language processing, speech analysis or any type of signal analysis.
These domains especially benefit from the recent developments in vision-based systems (VBS) due to the wide availability of training data and their relevance for many downstream applications.
DL models use deep multi-layer neural networks as a model to learn complex tasks from raw data, including the relevant features in this data.

\begin{figure}
    \centering
    \subfloat[Traditional software development]{\includegraphics[width=0.4\columnwidth]{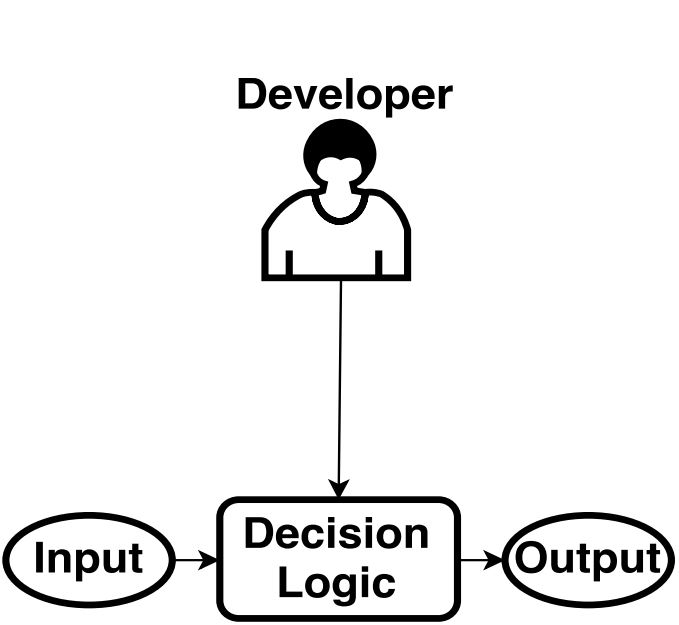}}\hfill
    \subfloat[Development for DL models]{\includegraphics[width=0.55\columnwidth]{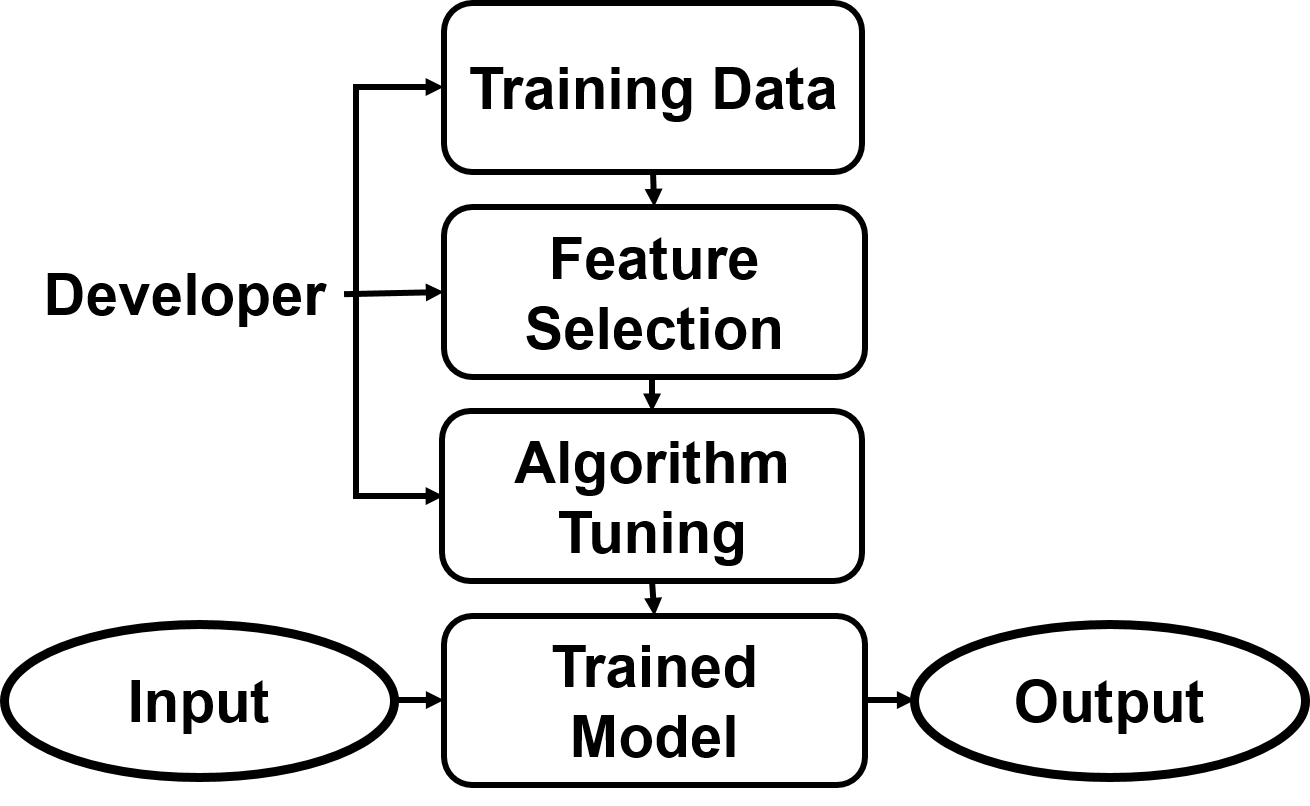}}
    \caption{Comparison of software development methods.}
    \label{fig:Comparison_Soft_Dev}
\end{figure}

As shown in Fig.\ref{fig:Comparison_Soft_Dev}, the construction of DL models has differences to the way software is traditionally developed~\cite{Breck2017,Amershi2019}. %
Instead of implementing the exact system behaviour, these models are derived from a dataset and the result is practically a black-box that cannot be analyzed and explained to the same degree as source code \cite{Numan2020}.
Nevertheless, DL models are of course software artifacts. 
It is important to note that DL models are usually not deployed as standalone systems, but are integrated with other software components and services, leading to another set of software artifacts within the overall system.
Accordingly, established best practices of software engineering apply to the development of DL models, too, such as continuous integration, version control, reusable and organized source, and, leading to the focus of this article, software testing~\cite{Breck2017,Amershi2019,Braiek2020}.

Still, DL models require the same degree of testing like any other software component to ensure they fulfill their functional and non-functional specifications. It has been observed that traditional software testing methods such as boundary value analysis and partition testing can be used to test DL models \cite{Herbold2020} but, in many cases, the testing of DL models only considers the performance on the test set, a set of previously unseen inputs and their correct outputs, using a set of evaluation metrics for the specific task.
Common metrics include the accuracy in classification, i.e. how many samples of the test set were assigned the correct class label or the mean-squared error in regression tasks where continuous values are predicted.
Due to the unique characteristics of DL models, exhaustive testing of these models is not achievable though performance measures are required to pick a model for deployment.
However, these matrices alone are insufficient to be confident about the overall model quality.
Additional criteria involve the generalization behaviour for real-world data when deployed, its robustness, as well as fairness, potential biases and interpretability respectively explainability of the model.

\begin{figure}
    \centering
    \subfloat[Original: White Shark]{\includegraphics[width=0.49\columnwidth]{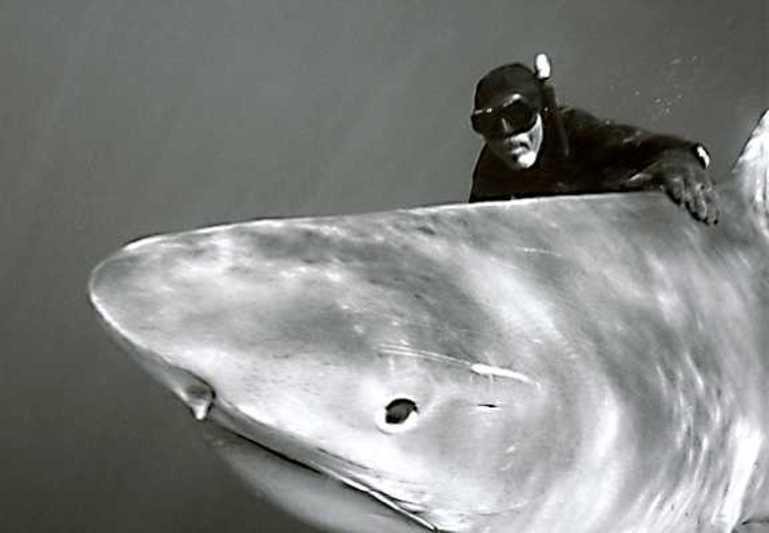}}\hfill
    \subfloat[Manipulated: Washbasin]{\includegraphics[width=0.49\columnwidth]{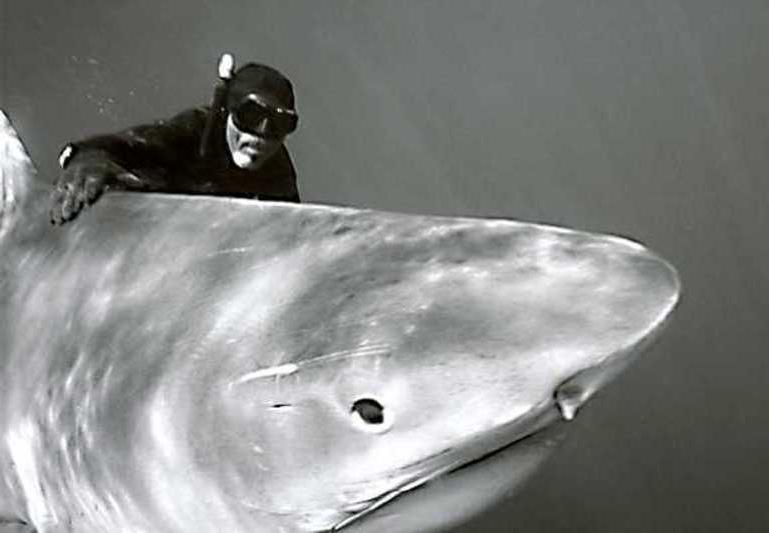}}
    \caption{A simple manipulation like horizontal flipping turns a correct classification (white shark) into a false classification (washbasin) in an image classification model, that is distributed via the model zoo of the popular PyTorch deep learning framework. Systematic testing of deep learning systems is necessary to identify weaknesses and increase robustness.}
    \label{fig:fool}
\end{figure}

While all of these criteria deserve dedicated attention, we focus our interest in this paper mostly on test data quality and the robustness of the model w.r.t. new or manipulated inputs.
Trivial or hidden changes in the input data can cause the model to change its predictions, such as shown in Fig.\ref{fig:fool}. 
Here, the change in the image should have no effect on the classification output as the content for the observer stays the same.
Nevertheless, the classification result is different and, obviously, wrong.
Evaluating these aspects is relevant for awareness of the limitations of the model, but also to gather information for improvements in subsequent development.
By training with more relevant training data, depending on the kind of weaknesses detected, the performance of the system will increase and the risk of failures will accordingly decrease.
At the same time, while testing a software system, one concern is often on reducing the testing effort, e.g. using only a limited set of data to accelerate testing~\cite{spieker2019towards}, and making best use of available resources for test execution.

This article reviews a selection of software testing methods that have shown to be especially suitable for testing DL models used in VBS.
It is not an exhaustive literature review, but highlights selected methods that have shown to be interesting and promising approaches, according to the authors' knowledge.
For an exhaustive and thorough review of testing machine learning, we refer to \cite{Zhang2020}.
Our discussion of testing is limited towards the validation of already trained models or models during training. We provide an initial experimental comparative analysis of different testing technique on a VBS-relevant case study. This experimental comparison allows us to draw initial conclusions on the necessity to combine testing techniques to strengthen the quality of trained DL models.

\section{Challenges in Testing DL Models}

Despite significant progress in the field of DL practices \cite{Amershi2019}, specific challenges exist for testing DL models.
These challenges can be broadly categorized as those of:
\begin{enumerate}
    \item Ensuring the quality of the model;
    \item Ensuring the quality of the training dataset;
    \item Predicting the exact output for a given input (a.k.a., the oracle problem);
    \item Defining test adequacy criteria given a vast input space;
    \item Protecting against adversarial attacks.
\end{enumerate}
In the following, We assume that an appropriate model architecture has been chosen and hyperparameters have been tuned from experimentation. 
We focus here on model training and test data selection, as well as adversarial detection.

\subsection{The Quality of the Model}
During model training, errors can arise from inaccurate user requirements, or from the inappropriate selection of model. In particular, the choice of an inappropriate network architecture, i.e., non-fitting network size, can render unsuitable the model in its learning task.  
These errors are usually addressed by requirements engineering through both application and DL experts to understand the problem and how it translates to DL terminology and methodology.
Furthermore, the hyperparameters for training need to be carefully chosen to lead to an efficient model.
This issue is an aspect of ongoing research and the selection of the right hyperparameters for training can make the difference between a mediocre and a state-of-the-art model \cite{Herbold2020}.
Common approaches are to start from a set of parameters used in research publications and manually adjust them until good performance is reached or to systematically search for a good set of hyperparameters using optimization techniques, such as random or grid search, or Bayesian optimization.
A systematic evaluation of different hyperparameters is costly as it requires the repeated training of the model, but it also gives higher confidence in the quality of the model and its parameters.

\subsection{The Quality of Training Data}
The behavior of DL models follows the examples given in the training data. 
When collecting a training dataset for a new application, it is important that the data is diverse enough to cover all variances that are expected to be encountered by the deployed model, i.e. the data distribution for which and on which the model is trained.
There are multiple potential errors here.
First, the collected training data might be insufficient to train a model that can generalize to the real-world data.
In this case, the training data should be expanded and the test set needs to be enriched by examples that the model does not generalize to. 

Second, the training data might lead to a capable model for the real-world data, but over time the data encountered by the deployed model gradually changes and no longer fits the initial training distribution.
This reduces the model performance and increases false predictions.
An approach to address this challenge is to monitor the actual data encountered by the deployed model and measure how it differs from the training data.
This monitoring is then followed by frequent model fine-tuning or retraining with an adjusted data set.
Other failures are caused by not carefully selecting the data and thereby introducing, for example, redundant data, adversarial data, or biases into the dataset.
In case of biases, these can lead to a model that does not make fair predictions, but is conditioned to repeat the biases encoded in the training data.

\subsection{Test Input and Output Selection}
DL models typically operate in application areas dealing with a large amount of data.
This creates a vast test input space, and consequently a significant challenge of selecting an adequate set of inputs. 
Answering to the question "\textit{when to stop testing}" leads to the definition of specific coverage criteria for DL models, which can guide the selection of inputs able to reveal false classification or incorrect regression during testing.
This problem has been encountered in the software testing domain before, especially in the area of fuzzing \cite{Manes2019}.

Of further relevance is the \textit{oracle problem}~\cite{Barr2015}.
For systems with smaller input space, test cases can be specified with specific inputs to the system and expected outputs for these inputs, known as \textit{test oracles}. 
However, as the number of allowable inputs is practically nearly infinite, it is practically impossible to exhaustively specify the expected output for all possible inputs against which the actual output can be compared in testing.
Therefore the correctness of the output in testing DL models cannot be easily determined.

\subsection{Adversarial Detection}
DL models are shown to be vulnerable to input artifacts purposely designed to attack the DL model, called \textit{adversarial inputs}, and and its security and privacy~\cite{Papernot2018}. %
Adversarial inputs can have different goals, such as forcing a misclassification of an input, to extract information about the model architecture, or the data used to train it, which might be personal information \cite{Papernot2018}.
Other attacks poison the training data and cause the model to learn the wrong behaviour or overload the model to reduce its capacity.

Adversarial inputs are not unique to computer vision models, but in this domain the effect is especially well investigated.
Defenses against adversarial inputs that make the models more robust as well as attacks to find new ways to generate adversarial inputs are an area of active research.
Currently, there is no definite way to avoid or detect all adversarial inputs in neural networks and for every new defense, an attack that circumvents it is published shortly after. 
To handle and address these selected challenges of testing DL models, it is crucial to adapt established methods from software testing for the testing of DL models.

\subsection{Fault Sources in DL models}
DL models are affected from the same fault sources as other software systems, such as incomplete specification or implementation errors, but, because of being trained from data, an additional set of fault sources has to be considered. Errors can occur in the selection of data used to create training datasets. We distinguish in the following biased inputs from adversarial inputs.

\paragraph{Biased Inputs}
Datasets are affected by many potential biases. Even though the problem of biases in datasets is largelly documented for personal data \cite{Roselli2019}, only few results are known regarding to biases in data used in VBS \cite{Luo2021}. It is worth noticing that industrial datasets are usually created by monitoring the execution of VBS in their execution environment and historical biases tend to be silently incorporated into the datasets which are then used for training DL models. Detecting and removing these biases is an active research area known as {\it anomaly detection}, unfortunately not yet sufficiently focused on detecting errors in industrial datasets for VBS.

\paragraph{Adversarial Inputs}
Adversarial inputs address a structural weakness of DL methods. By generalizing behaviors from training datasets, DL models are susceptible for manipulated inputs. These manipulations can generate adversarial inputs that mislead the DL model and falsify the predictions. Adversarial inputs are not specific to DL models, their effect is especially strong in this case as the changes are possible without being noticeable by human observers or controlling methods. 
When deploying software testing for DL trained models, there are actually two main issues: i) How to predict the expected output of a given test input (a.k.a., \textit{the oracle problem})? ii) How to select adequate test inputs in order to increase confidence in the model correctness (a.k.a., \textit{the test adequacy problem})?

\section{Deep Learning Testing Techniques}
This section presents test adequacy criteria specific to DL and reviews four potential techniques for testing DL models. 

\subsection{Test Adequacy Criteria for DL}
To address the challenge mentioned above, test adequacy criteria have been proposed for DL.
Inspired by traditional code coverage metrics such as statement or decision coverage, \textit{neuron coverage} \cite{pei2017deepxplore} counts the number of activated neurons in the neural network when test inputs are submitted for classification. 
The study reports that the higher the neuron activation coverage, the higher the chances of detecting wrong classifications. 
Other studies have confirmed this initial result and also extended the proposed structural criteria by distinguishing neuron-level criteria from layer-level coverage criteria \cite{ma2018deepgauge}. 
Instead of looking at the network structure, a different criteria called \textit{surprise adequacy} has been introduced for measuring the diversity in training data \cite{Kim2019}. 
The surprise of an input is defined as the difference between the input and the training dataset w.r.t. the behavior of the DL model. 
It is then suggested to select inputs which are sufficiently surprising but still within the expected data distribution to compose an adequate test set.
These criteria are specific to DL and can be used to guide the selection of test inputs, but techniques are needed to evaluate the correctness of results when test inputs are submitted to DL models, which we will discuss in the following.

\subsection{Differential Testing (DT)}
In Differential Testing (DT) the functionality of a system is assessed by comparing the behaviour of multiple implementations or models for the same task.
All these systems are executed with the same inputs and the difference between their results is used to assess the correctness of the tested program.%
DT's main issues are 1) how to identify the faulty system, as there is no way to determine which system provides the correct answer and 2) how to exploit a test set of sufficient quality to get confidence in the correctness of the tested system.  %

\emph{DeepXplore} \cite{pei2017deepxplore} was the first method to explore DT to test DL models.
By using the publicly-available datasets MNIST (dataset for hand-written digits), ImageNet (dataset for general image classification), VirusTotal (dataset for malicious PDF documents), Udacity video dataset for driving, and Drebin (dataset for Android apps), DeepXplore tested state-of-the-art DL models and showed that neuron activation coverage coupled with DT can produce inputs depicting incorrect behaviors by selecting those inputs which lead to wrong classification for most models. 
Interestingly, these inputs can be used to re-instruct the tested model to enhance its accuracy.
However, the accuracy attained for the MNIST dataset after employing the approach presented in DeepXplore was 98.96\%, which is 0.95\% less than the state-of-the-art model.

Whereas DeepXplore generates inputs via gradient ascent optimization, Chunyan Wang \cite{Wang2019a} proposed differential combination testing based on the selection of transformation functions in addition to the evaluation via DT.

\subsection{Metamorphic Testing (MT)}

Metamorphic Testing (MT) is an effective method to deal with programs for which an oracle is difficult or even impossible to define. In MT, 
known relations between the inputs and outputs of a program, called metamorphic relations (MR), can be used to find faults in the program \cite{Segura2018a}. 
For example, in Figure~\ref{fig:fool}, a typical MR says that applying horizontal flipping over the image should not change its classification label, even though the correct label is unknown.
Starting from an initial source test case, by using MRs, it is possible to generate follow-up test cases for which the expected result has to satisfy properties defined by the MRs.
If the relation is violated, then a fault is revealed, even though we ignore where is the fault located and what is the correct expected output of the program.
While an exact test oracle is usually impossible to get for DL, due to the stochastic nature of the predictions computed by the model, many MRs can often easily be identified and conjointly used to check the model outputs.
As an example, consider an object detection system that detects a number of objects in an image, computes surrounding boxes on these objects and classifies them. For such a system, several MRs can be defined. For instance, flipping the images left-right or upside-down, shall lead obviously to the same label of the image, but the coordinates of computed surrounding boxes can also be easily predicted in the transformed image. Similarly, if the image is transformed by adding objects without overlapping or changing existing objects, then a relation can be established to verify that the number of detected objects must be greater than the number in the original image.
Note that even though a difference between the results certainly reveals a fault, nothing guarantees that at least one of the two results is correct. 
In that sense, MT suffers from the same issue as DT.

\begin{table*}
	\centering
	\caption{MT of DL models using different MRs like rotation with different angles, different width and height settings of the images, modifying the shearing and changing the zooming of the images.}
	\label{tab:RT}
	\centering
	\begin{tabular}{|c|c|c|c|c|c|c|c|c|c|c|c|c|c|} \hline
	
	Relation                & \multicolumn{3}{|c|}{Rotation} & \multicolumn{4}{|c|}{Width and Height Shift} & \multicolumn{4}{|c|}{Shearing} & \multicolumn{2}{|c|}{Zooming}  \\ \hline
    Augmentation            &   30   &   60    & 90  &   0.1   &   0.2    & 0.25 & 0.5 &   25   &   45    & 65 & 85 &   [0.5,1.5]  &   [2.5,3.5]  \\ \hline

    Without Aug             & \multicolumn{13}{|c|}{99.26}        \\ \hline
    Only Train Aug          & 99.03  &  99.06  & 98.72 & 99.28  & 98.96   & 98.9 & 98.14 & 99.3  & 99.23   & 99.27 & 98.91   &  99.26 & 17.21    \\ \hline
    Only Test Aug           & 96.23 & 74.99  &  57.04 & 93.02  &  56.11  & 40.67 & 17  & 98.1  & 91.63   & 76.95 & 62.48 &   90.24 & 23.56   \\ \hline
    Train and Test Aug      & 99.2  & 98.8   & 98.45 & 99.16  & 98.64   & 98.24 & 86.82 & 99.13  & 98.92   & 98.66 & 97.27 &  98.81 & 97.18     \\ \hline
	\end{tabular}
\end{table*}

\begin{table}[t]
	\centering
	\caption{MT using two different combinations of MRs: rotation and weight and height shift.} 
	\label{tab:RWHT}
	\centering
	\begin{tabular}{|c|c|c|} \hline
    Rotation     &   30  &   60     \\
    Width and Height Shift     &  0.1  &   0.2     \\ \hline

    Without Aug          & \multicolumn{2}{|c|}{99.26}        \\ \hline
    Only Train Aug          & 87.33  & 39.07         \\ \hline
    Only Test Aug           & 99.32  & 98.38       \\ \hline
    Train and Test Aug       & 98.74 & 96.79          \\ \hline
	\end{tabular}
\end{table}

\begin{table}[t]
	\centering
	\caption{MT using three different combinations of MRs: rotation, and width and height shift, and shear.}
	\label{tab:RWHST}
	\centering
	\begin{tabular}{|c|c|c|} \hline
    Rotation     &   30  &   60     \\
    Width and Height Shift     &  0.1  &   0.2     \\ 
    Shear     &  25  &   45     \\ \hline

    Without Aug          & \multicolumn{2}{|c|}{99.26}        \\ \hline
    Only Train Aug          & 82.85  & 34.04        \\ \hline
    Only Test Aug           & 99.32  & 98.08        \\ \hline
    Train and Test Aug       & 98.46  & 93.72      \\ \hline
	\end{tabular}
\end{table}

\begin{table}[t]
	\centering
	\caption{MT with four different settings of MRs: rotation, width and height shift, shear, and zoom.} 
	\label{tab:RWHSZT}
	\centering
	\begin{tabular}{|c|c|c|} \hline
    Rotation     &   30  &   60     \\
    Width and Height Shift     &  0.1  &   0.2     \\ 
    Shear     &  25  &   45     \\ 
    Zoom     &  [0.5, 1.5]  &   [2.5, 3.5]     \\ \hline

    Without Aug          & \multicolumn{2}{|c|}{99.26}        \\ \hline
    Only Train Aug          &  68.43 & 17.99        \\ \hline
    Only Test Aug           &  35.09 & 36.01       \\ \hline
    Train and Test Aug       & 96.52  & 90.5    \\ \hline
	\end{tabular}
\end{table}

Using MT to test learning models originates from \cite{murphy2008properties}, but only recent works have pushed MT to validate DL models by using dedicated metamorphic relations \cite{ding2017validating,dwarakanath2018identifying,Zhang2018}. 
The proposed approaches have been used for identifying implementation bugs in ML based image classifiers, validating other DL frameworks used in diverse applications, and validating a DL classifier for automated organization of biological cell images.
MT is relevant for testing DL frameworks as the training algorithms are complex and the training of DL models require huge amount of data. 
However, the identification of MRs, that is the most beneficial relations for testing a system, is challenging. 
In \cite{dwarakanath2018identifying}, many relations for image classifiers are proposed and experiments show that some of these metamorphic relations are able to identify 71\% of the implementation bugs in average. This level of accuracy appear to be largely insufficient for considering the deployment if these models in safety-critical VBS.
Recently, it was further proposed to use contextual bandits to select the most appropriate MR for image classification and object detection \cite{spieker2019adaptive}. The attempt was to learn how to adapt the selection of the most promising MR for testing DL models. Even though, the approach of \cite{spieker2019adaptive} was not deployed to test DL models embedded into VBS, the method reveals itself very powerful to select appropriate MRs. 

On another research lead, the testing tool \emph{DeepTest} \cite{tian2018deeptest} was proposed to detect inaccurate behavior of self-driving cars.
Using data sources coming from dedicated sensors (e.g., LiDAR or camera), decision systems of self-driving cars have to be thoroughly tested in diverse driving conditions, because inaccurate behaviour might cause fatal collisions.
DeepTest leverages neuron activation coverage and MT to create synthetic test images which represent realistic weather or lighting conditions.
For example, for a given driving situation, the self-driving car’s steering angle should not change under any lighting-varying conditions. 
\cite{tian2018deeptest} suggests that MT is appropriate for testing DL-based decision systems of autonomous vehicles. Besides these work, little is still known on the deployment of MT to test DL models in VBS. It is a promising research area that has its own challenges.

Inspired from DeepTest, we designed some experiments to see how MT can be used to test trained DL models on MNIST dataset. 
However, we didn't refer the DeepTest code because the method's overall outcome was to forecast steering angle for autonomous automobiles rather than classification.
As a result, we developed four MRs based on various rotations, image height and width changes, image shearing, and image zooming angles modifications.
We conducted tests in three distinct approaches, the first of which included modifying the MRs separately (results shown in Table \ref{tab:RT}).
Second, merging two MRs (rotation, and width and height shift) and examining the results (as shown in Table \ref{tab:RWHT}).
Third, combine and tune three separate relations (rotation, width and height shift, and shearing), as shown in Table \ref{tab:RWHST}.
Finally, we used all four relations (rotation, width and height shift, shearing, and zooming) and tuned them to observe how they affect the final accuracy, as shown in Table \ref{tab:RWHSZT}.

We ran four types of tests to examine how these MRs impact the model.
First, we look at the model's accuracy on the test dataset without any train or test data augmentation (represented as Without Aug in Table \ref{tab:RT}-\ref{tab:RWHSZT}). 
Second, we ran tests to examine how the model responds to augmented training data while testing it on non-augmented test data (represented as only Train Aug in Table \ref{tab:RT}-\ref{tab:RWHSZT}).
Third, we used original training data to train the model but then used augmented data to test it (represented as only Test Aug in Table \ref{tab:RT}-\ref{tab:RWHSZT}).
Finally, we tested DL models that had been trained on augmented training data and tested them with augmented test data (represented as Train and Test Aug in Table \ref{tab:RT}-\ref{tab:RWHSZT}). 

Based on the acquired results, we discovered that the DL model works effectively when trained with either original or augmented data. 
However, if the DL model is not trained with augmented data but tested on augmented data, it struggles to predict on the test set.
In another example, we can observe that when the DL model is trained and evaluated using augmented data, it performs significantly better.

\subsection{Mutation Testing (MuT)}

Mutation Testing (MuT) is a traditional testing method, used to assess the quality of a test set and to generate fault-revealing test cases~\cite{Papadakis2018a}.
In MuT, mutants are modified versions of the program under test and the goal is to evaluate a test suite by its ability to distinguish the original program from its mutants.
A mutant is generated by applying one or multiple mutation operators on the original program, i.e., by only introducing minor changes in the data set or the model architecture~\cite{Jahangirova2020}.
When a test case observes a difference in the behaviour of the original program and its mutant, we say that it "kills" the mutant.

MuT was deployed for testing DL models in \cite{Zhu2018DeepMutationMT} with a tool called \emph{DeepMutation}.
The authors present both source-level mutation operators, that manipulate training data and code, as well as model-level mutation operators, which manipulate the already trained model.
DeepMutation can thereby produce mutants that allow to evaluate the quality of a test set.

We ran several tests on the MNIST dataset with the code provided by \cite{Zhu2018DeepMutationMT} to assess how well DeepMutation evaluates the test set.
To examine how many test samples could be misclassified, we utilized model-level mutation operators with varied mutation ratios, the results of which are reported in Table \ref{tab:MLGF}, and \ref{tab:MLNEB}. 

\begin{table*}
	\centering
	\caption{Mut of DL models using different model-level tuning like Gaussian fuzzing, weight shuffling, neuron effect block, neuron activation inverse, and neuron switch.}
	\label{tab:MLGF}
	\centering
	\begin{tabular}{|c|c|c|c|c|c|c|c|c|c|c|} \hline
	Type               & \multicolumn{2}{|c|}{Gaussian Fuzzing} & \multicolumn{2}{|c|}{Weight Shuffling} & \multicolumn{2}{|c|}{Neuron Effect Block} & \multicolumn{2}{|c|}{Neuron Activation Inverse} & \multicolumn{2}{|c|}{Neuron Switch} \\\hline
    Mutation Ratio     &   Accuracy  &   Error  &   Accuracy  &   Error &   Accuracy  &   Error  &   Accuracy   &   Error  &   Accuracy &   Error     \\\hline
     0.01              &     96.39   &   3.61   &    96.97    &   3.03  &   96.92     &   3.08   &    96.97     &   3.03   &    96.95   &   3.05 \\ \hline
     0.1               &     90.1    &   9.9    &    66.11    &   33.89 &   79.56     &   20.44  &    32.04     &   67.96  &    94.63   &   5.37\\\hline
     0.2               &     85.06   &   14.94  &    32.32    &   67.68 &   59.3      &   40.7   &    18.41     &   81.59  &    67.82   &   32.18 \\ \hline
     0.3               &     82.71   &   17.29  &    30.33    &   69.67 &   49.04     &   50.96  &    7.22      &   92.78  &    35.22   &   64.78 \\ \hline
     0.4               &     58.43   &   41.57  &    16.18    &   83.82 &   16.83     &   83.17  &    6.64      &   93.36  &    18.67   &   81.33\\\hline
     0.5               &     33.17   &   66.83  &    14.64    &   85.36 &   11.12     &   88.88  &    6.97      &   93.03  &    14.86   &   85.14\\\hline
	\end{tabular}
\end{table*}

\begin{table}[t]
	\centering
	\caption{Mut of DL models using different model-level tuning.}
	\label{tab:MLNEB}
	\centering
	\begin{tabular}{|c|c|c|} \hline
    Type                            &   Accuracy      &     Error     \\\hline
     Layer Deactivation             &    7.17         &     92.83     \\ \hline
     Layer Addition                 &    0.07         &     99.93   \\\hline
     Activation Function Removal    &    87.9         &     12.1    \\ \hline
	\end{tabular}
\end{table}

In another study \cite{Wang2019c}, MuT is applied as a defense mechanism for the detection of adversarial examples.
The authors report that adversarial inputs are more sensitive to mutations of the DL model than the original inputs.
By exploiting this observation, their method mutates the DL model and from the resulting changes in the output, it is determined whether an input is likely to be adversarial.
The evaluated analysis determine the quality of the test data and the level to which the introduced errors could be detected.

\subsection{Combinatorial Testing (CT)}

Combinatorial Testing (CT) is a software testing approach that uses value interactions coverage of input data to measure the quality of test suites and evaluate software systems~\cite{kuhn2009combinatorial}. 
CT follows the assumption that input parameters often interact with each other and certain tuples of values may actually lead to faults~\cite{kuhn2004software}.
Focusing on dedicated interactions between parameters (typically pairs or triple of values) on the basis of representative values instead of performing an exhaustive search of the whole input space allows thereby a more structured but also more efficient testing of the system.
This efficiency advantage has also been a motivating factor for the application of CT in DL systems.
Here, CT is considered to be a black-box testing technique which delves into the network to co-relate between architectural neurons. CT is of two types; fixed strength and variable strength CT both of which have been explored in \cite{ma2019deepct}, and \cite{chen2019variable}.

In \cite{ma2019deepct}, the authors have used CT in order to evaluate the system robustness by reducing the spaces of runtime states.
After each layer, their technique (DeepCT) analyzes the interactions between the neurons. 
The technique is useful in testing the large runtime states specifically neuron outputs in DL models. 
They proposed CT coverage guided test generation technique which was useful in detecting the defects at early stages.
Furthermore, Lei Ma  \cite{chen2019variable} explored the variable strength CT to test how each neuron present in the post layer is influenced by the combinatorial effect of all neurons present in pre-layer. 
To accomplish this, three different methods are introduced to establish variable strength CT interaction patterns for DNNs.

\subsubsection{Combinatorial Smoke Testing}
Smoke Testing (ST) is a degree of testing performed by developers on the first software builds to ensure that all of the basic/core features are operating properly. Due to a number of issues like configuration, regression, code, and environmental issues, the environment is not well served by the software builds. So, after the development team has completed the initial build, it is subjected to basic testing before being submitted to the next stage of testing. This initial level of testing is known as smoke testing. Using this testing technique, the authors of \cite{herbold2020smoke} performed experiments on both supervised (classification) and unsupervised (clustering) learning and they came up with universal smoke tests using techniques like boundary value analysis, and equivalence classes, which verifies the fundamental functions that can be performed without failing.
\begin{figure*}
    \centering
    \subfloat[Differential Testing]{\includegraphics[height=5cm]{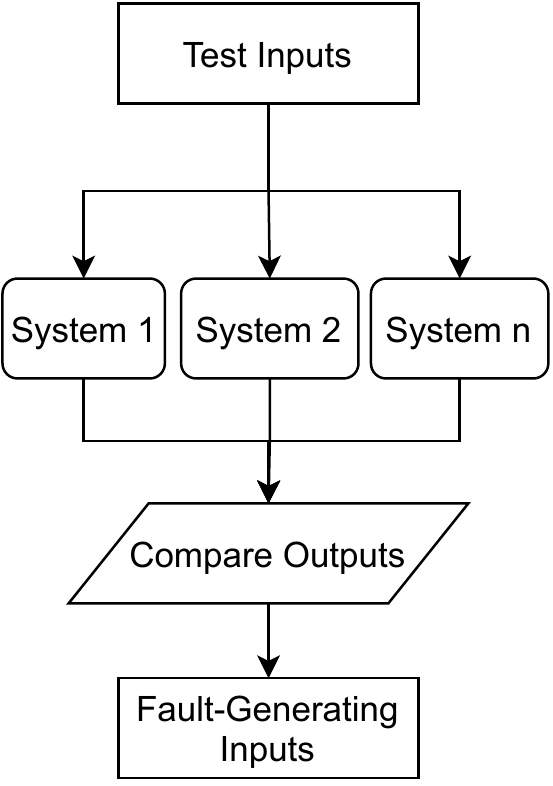}}\hfill
    \subfloat[Metamor. Test.]{\includegraphics[height=5cm]{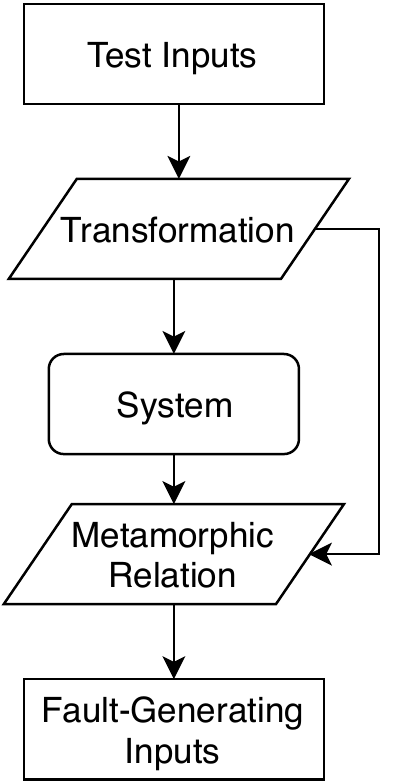}}\hfill
    \subfloat[Mutation Testing]{\includegraphics[height=5cm]{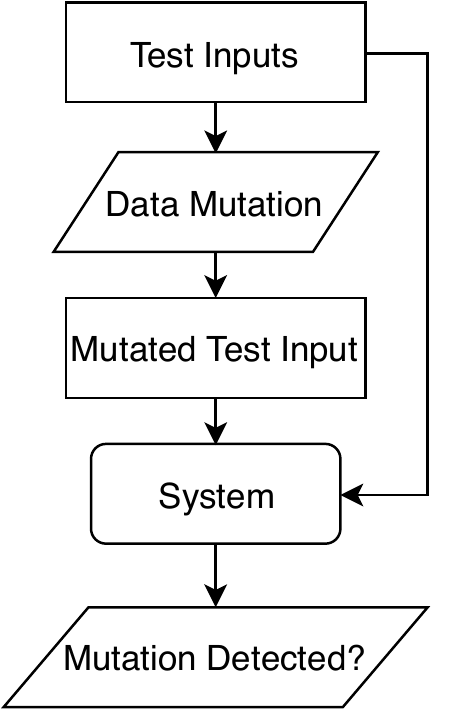}}\hfill
    \subfloat[Adversar. Pert. Test.]{\includegraphics[height=5cm]{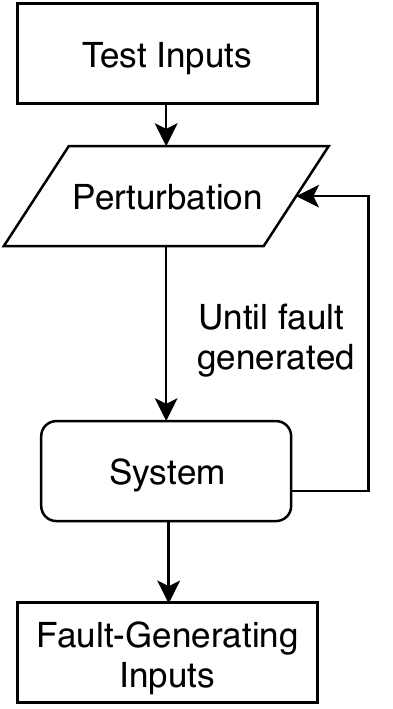}}\hfill
    \subfloat[Combinat. Test.]{\includegraphics[height=5cm]{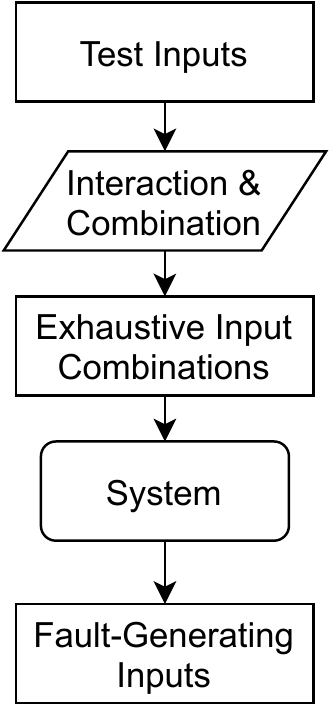}}
    \caption{Schematic Overview of Testing Techniques for Deep Learning Systems}
    \label{fig:dl_testing_techniques}
\end{figure*}
\subsection{Adversarial Perturbation Testing (APT)}
A technique developed to attack machine learning systems, known as \emph{adversarial perturbation} can be tuned to serve the purpose of testing DL models. 
An adversarial example is an intentional attack created to fool a DL model with a test input which makes the system produce a wrong answer \cite{Papernot2018}. 
Interestingly, adversarial inputs are close to correctly classified inputs, so that they remain undetectable for human eyes or classical detection means.
Adversarial Examples can be referred as an optical illusion but for machines (see Figure \ref{fig:fool}).
Szegedy et al. \cite{szegedy2013intriguing}, were one of the first to point out the existence of adversarial examples. 
They pointed that given an input (x) and a target (t), which comes out to be C(x) $\neq$ t, it is often possible to find similar inputs (x) which can result in C(x) = t yet both the inputs (x and x) are close according to some distance metric. 
This property is known as a targeted adversarial example. 
There are other less powerful attacks available in literature like untargeted adversarial examples which focuses on classifying (x) as any wrong target class.
Nguyen et al. \cite{nguyen2015deep} have shown that an image can be certainly changed in a manner that the changes are unnoticeable to human eyes. 
Those changed images can cause the deep neural network to mislabel the image. It shows that it is easy to generate images unnoticeable to human (fooling images). 
However the neural network confidently classify it as a recognizable object with 99.9\%. 
They have taken CNN networks to perform the perturbation on MNIST data sets or ImageNet, further finding images with algorithms or gradient ascent due to which DNN labels the image with high confidence as a member of recognizable group. 
They have shown that DNN models are fooled easily, to categorize unrecognizable images with high probabilities as a member of a recognizable group. 
The DNN considers the objects as nearly perfect samples of recognizable images. 
By generating adversarial examples, Adversarial Perturbation Testing (APT) can produce test cases which reveal wrongly classified inputs, by construction. 
For instance, to evaluate the robustness of DL models, \emph{CleverHans} \cite{papernot2016technical} is a state-of-the-art open source toolkit to automatically generate adversarial inputs for machine learning models.
Where the authors have suggested a new algorithm to calculate the perturbations which efficiently fool DNN networks. 
This further helps to quantify the robustness of deep classifiers. 
DeepFool is built on the classifier being iteratively linearized to provide minimum perturbations that are adequate enough to modify classification labels. 
They did experiments on 3 data-sets and 8 classifiers, which showed the effective positive results in calculating the adversarial perturbations. 
The tool can calculate the least perturbed adversarial image and it can thus be converted into a APT tool, i.e., a method which generates test inputs able to push the DL model under test to produce wrongly classified images. 
The approach can be further refined by using Generative Adversarial Networks (GANs) which use a discriminating network to evaluate inputs produced by a generative network. 
By learning how to produce adversarial examples, the discriminating network is by construction an APT tool able to produce test cases which lead to wrong classifications for the tested DL model. 
An example is given by the \emph{DeepRoad} approach~\cite{Zhang2018}. 
Here, GANs are combined with MT to produce adversarial examples for testing decision systems of self-driving cars.
However, it is worth noting that GANs are difficult to train in practice with possible non-convergence issues, which has the side-effect to increase the complexity of the testing setup.
In \cite{papernot2016limitations}, they have presented a unique class of algorithms by formalizing the space of adversaries in neural networks. 
This novel class of algorithms has been introduced to craft the adversarial examples on the basis of accurate mapping understanding amongst DNNs input and output. 
When applied to a computer vision, their algorithm was able to produce the samples for human subjects, with 97\% adversarial success rate, by modifying the input features on the average of only 4.02\% per sample. 
They have shown the vulnerability of various samples to adversarial perturbations. 
They have also shown the vulnerability of various samples to adversarial perturbations by introduction of hardness measure.

\section{Discussion}%
Generally speaking, it can be concluded from our initial experiments that MuT can be used to test both training and test data. MT, on the other hand, can be used to assess the correctness of machine learning models.
DT can be used to select inputs leading to inaccurate classification and use those inputs to enhance the model accuracy. However, in cases where testing the DL models is time-constrained, MuT can be utilized to identify the model's corner cases.
Fig.\ref{fig:dl_testing_techniques} gives a schematic overview over all the presented testing techniques and their general workflow. As seen above, these techniques brings specific challenges when they are deployed to test DL models. Nevertheless they are complementary in their ability to cover the issues raised by the deployment of DL models.
Tab.\ref{tab:CT} summarizes how testing techniques address the challenges of testing DL models.
\begin{table}[t]
	\centering
	\caption{Testing Techniques vs. Challenges}
	\label{tab:CT}
	\begin{tabular}{|c|c|c|c|c|c|} \hline
		                 & DT    & MT    & MuT  & APT   & CT    \\ \hline
Quality of the Model     &	     &  \cm  &  \cm &  \cm  &  \cm     \\ \hline
Quality of Training Data &       &       &  \cm &  \cm  &    \\ \hline
Oracle Problem           &  \cm  & \cm   &      &       &       \\ \hline
Test Input Selection     &       & \cm   &  \cm &       &  \cm \\ \hline
Adversarial Detection    &       &       &      &  \cm  &       \\ \hline
	\end{tabular}
\end{table}
Model quality is addressed by MT, MuT, APT, and CT as these techniques stress the model and evaluate its correctness and performance. However, the training dataset quality is addressed exclusively by MuT and APT as only those techniques can modify or analyse the network structure.
Conversely, only DT and MT tackle the oracle problem as the other techniques exclusively exploit the tested DL model by counting exclusively on an external oracle to evaluate the test results. 
Test input selection is addressed by MT, CT, and MuT because these techniques can generate systematic fault-revealing test cases. To some extent,  APT has also some test inputs generation that but not at a great scale. 
Finally, only APT deals with adversarial detection because the technique itself is based on the creation of adversarial examples. 
In summary, Table \ref{tab:CT} shows that these testing techniques for DL are complementary and should be combined based on the peculiar targeted application. For instance,  the combination of MT and APT, or DT, MuT and APT, offers us an ideal coverage of the five identified problems. 
Depending on the deployment context of the VBS, it can be the combination of several of these testing techniques that is the most relevant.

\section{Conclusion}
In this paper, we have presented techniques for testing DL models. 
We have seen that challenges such that quality of the model, quality of the training dataset, the oracle problem, the test input selection and adversarial detection can be effectively tackled by a combination of testing methods. 
Our conclusion is that methods from the traditional software testing area, when they are combined altogether, are suitable for the testing of DL models even though they bring specific challenges that need to be handled. 
In particular, differential testing, metamorphic testing and adversarial perturbation testing attract interest due to their ability to mitigate the test oracle problem, e.g. by defining relations between source test cases and generated test cases, and their ability to detect adversarial attacks. 
However, the identification and selection of the most fault-revealing metamorphic relations is challenging for image classification and object detection systems.
Mutation testing, on the other hand, allows one to focus on the consistency of a model towards variants of similar inputs. 
Its deployment for testing DL models is appealing even though the definition of mutation operators is not trivial.  
In the context of testing DL models, it can be used to analyze strengths and weaknesses of both the model and the datasets, without the manual creation of additional test cases. 
Combinatorial testing for DL models is interesting to evaluate the model robustness and adversarial perturbation testing evaluates security aspects of DL models.

\bibliographystyle{IEEEtran}
\bibliography{refs}

\begin{thebibliography}{10}
\providecommand{\url}[1]{#1}
\csname url@samestyle\endcsname
\providecommand{\newblock}{\relax}
\providecommand{\bibinfo}[2]{#2}
\providecommand{\BIBentrySTDinterwordspacing}{\spaceskip=0pt\relax}
\providecommand{\BIBentryALTinterwordstretchfactor}{4}
\providecommand{\BIBentryALTinterwordspacing}{\spaceskip=\fontdimen2\font plus
\BIBentryALTinterwordstretchfactor\fontdimen3\font minus
  \fontdimen4\font\relax}
\providecommand{\BIBforeignlanguage}[2]{{%
\expandafter\ifx\csname l@#1\endcsname\relax
\typeout{** WARNING: IEEEtran.bst: No hyphenation pattern has been}%
\typeout{** loaded for the language `#1'. Using the pattern for}%
\typeout{** the default language instead.}%
\else
\language=\csname l@#1\endcsname
\fi
#2}}
\providecommand{\BIBdecl}{\relax}
\BIBdecl

\bibitem{Breck2017}
E.~Breck, S.~Cai, E.~Nielsen, M.~Salib, and D.~Sculley, ``The {{ML}} test
  score: {{A}} rubric for {{ML}} production readiness and technical debt
  reduction,'' in \emph{{{IEEE International Conference}} on {{Big Data}}
  ({{Big Data}})}, vol.~47, 2017, pp. 1123--1132.

\bibitem{Amershi2019}
\BIBentryALTinterwordspacing
S.~Amershi, A.~Begel, C.~Bird, R.~DeLine, H.~Gall, E.~Kamar, N.~Nagappan,
  B.~Nushi, and T.~Zimmermann, ``Software engineering for machine learning: A
  case study,'' in \emph{Proceedings of the 41st International Conference on
  Software Engineering: Software Engineering in Practice}, ser. ICSE-SEIP
  '19.\hskip 1em plus 0.5em minus 0.4em\relax IEEE Press, 2019, p. 291–300.
  [Online]. Available: \url{https://doi.org/10.1109/ICSE-SEIP.2019.00042}
\BIBentrySTDinterwordspacing

\bibitem{Numan2020}
G.~Numan, ``\BIBforeignlanguage{en}{Testing {{Artificial Intelligence}}},'' in
  \emph{\BIBforeignlanguage{en}{The {{Future}} of {{Software Quality
  Assurance}}}}, S.~Goericke, Ed.\hskip 1em plus 0.5em minus 0.4em\relax
  {Cham}: {Springer International Publishing}, 2020, pp. 123--136.

\bibitem{Braiek2020}
H.~B. Braiek and F.~Khomh, ``\BIBforeignlanguage{en}{On testing machine
  learning programs},'' \emph{\BIBforeignlanguage{en}{Journal of Systems and
  Software}}, vol. 164, p. 110542, Jun. 2020.

\bibitem{Herbold2020}
S.~Herbold and T.~Haar, ``Smoke {{Testing}} for {{Machine Learning}}: {{Simple
  Tests}} to {{Discover Severe Defects}},'' \emph{arXiv:2009.01521 [cs]}, Sep.
  2020.

\bibitem{spieker2019towards}
H.~Spieker and A.~Gotlieb, ``Towards testing of deep learning systems with
  training set reduction,'' \emph{arXiv preprint arXiv:1901.04169}, 2019.

\bibitem{Zhang2020}
J.~M. Zhang, M.~Harman, L.~Ma, and Y.~Liu, ``Machine {{Learning Testing}}:
  {{Survey}}, {{Landscapes}} and {{Horizons}},'' \emph{IEEE Transactions on
  Software Engineering}, pp. 1--1, 2020.

\bibitem{Manes2019}
V.~J.~M. Man{\`e}s, H.~Han, C.~Han, S.~K. Cha, M.~Egele, E.~J. Schwartz, and
  M.~Woo, ``The {{Art}}, {{Science}}, and {{Engineering}} of {{Fuzzing}}: {{A
  Survey}},'' \emph{IEEE Transactions on Software Engineering}, 2019.

\bibitem{Barr2015}
E.~T. {Barr}, M.~{Harman}, P.~{McMinn}, M.~{Shahbaz}, and S.~{Yoo}, ``The
  oracle problem in software testing: A survey,'' \emph{IEEE Transactions on
  Software Engineering}, vol.~41, no.~5, pp. 507--525, 2015.

\bibitem{Papernot2018}
N.~Papernot, P.~McDaniel, A.~Sinha, and M.~P. Wellman, ``{{SoK}}: {{Security}}
  and {{Privacy}} in {{Machine Learning}},'' in \emph{2018 {{IEEE European
  Symposium}} on {{Security}} and {{Privacy}} ({{EuroS\&P}})}, 2018, pp.
  399--414.

\bibitem{Roselli2019}
D.~Roselli, J.~Matthews, and N.~Talagala, ``Managing bias in ai,'' in
  \emph{Companion Proc. of The 2019 World Wide Web Conf.}\hskip 1em plus 0.5em
  minus 0.4em\relax New York, NY, USA: ACM, 2019, p. 539–544.

\bibitem{Luo2021}
\BIBentryALTinterwordspacing
Y.~Luo, Y.~Xiao, L.~Cheng, G.~Peng, and D.~D. Yao, ``Deep learning-based
  anomaly detection in cyber-physical systems: Progress and opportunities,''
  \emph{ACM Comput. Surv.}, vol.~54, no.~5, May 2021. [Online]. Available:
  \url{https://doi.org/10.1145/3453155}
\BIBentrySTDinterwordspacing

\bibitem{pei2017deepxplore}
K.~Pei, Y.~Cao, J.~Yang, and S.~Jana, ``{DeepXplore}: Automated whitebox
  testing of deep learning systems,'' in \emph{Proceedings of the 26th
  Symposium on Operating Systems Principles (SOSP)}, 2017, pp. 1--18.

\bibitem{ma2018deepgauge}
L.~Ma, F.~Juefei-Xu, F.~Zhang, J.~Sun, M.~Xue, B.~Li, C.~Chen, T.~Su, L.~Li,
  Y.~Liu \emph{et~al.}, ``{DeepGauge}: Multi-granularity testing criteria for
  deep learning systems,'' in \emph{Proc. of the 33rd ACM/IEEE Int. Conf. on
  Automated Soft. Eng. (ASE)}, 2018, pp. 120--131.

\bibitem{Kim2019}
J.~Kim, R.~Feldt, and S.~Yoo, ``Guiding deep learning system testing using
  surprise adequacy,'' in \emph{Proceedings of the 41st International
  Conference on Software Engineering (ICSE)}, 2019, pp. 1039--1049.

\bibitem{Wang2019a}
C.~Wang, W.~Ge, X.~Li, and Z.~Feng, ``{{DCT}}: {{Differential Combination
  Testing}} of {{Deep Learning Systems}},'' in \emph{Art. {{Neural Networks}}
  and {{Machine Learning}} \textendash{} {{ICANN}}: {{Image Processing}}},
  2019, pp. 697--710.

\bibitem{Segura2018a}
S.~Segura, D.~Towey, Z.~Q. Zhou, and T.~Y. Chen, ``Metamorphic {{Testing}}:
  {{Testing}} the {{Untestable}},'' \emph{IEEE Software}, 2018.

\bibitem{murphy2008properties}
C.~Murphy, G.~Kaiser, L.~Hu, and L.~Wu, ``Properties of machine learning
  applications for use in metamorphic testing,'' in \emph{Proceedings of the
  Twentieth International Conference on Software Engineering \& Knowledge
  Engineering (SEKE)}, 2008, pp. 867--872.

\bibitem{ding2017validating}
J.~Ding, X.~Kang, and X.-H. Hu, ``Validating a deep learning framework by
  metamorphic testing,'' in \emph{IEEE/ACM 2nd International Workshop on
  Metamorphic Testing (MET)}, 2017, pp. 28--34.

\bibitem{dwarakanath2018identifying}
A.~Dwarakanath, M.~Ahuja, S.~Sikand, R.~M. Rao, R.~Bose, N.~Dubash, and
  S.~Podder, ``Identifying implementation bugs in machine learning based image
  classifiers using metamorphic testing,'' in \emph{Proceedings of the 27th ACM
  SIGSOFT International Symposium on Software Testing and Analysis (ISSTA)},
  2018, pp. 118--128.

\bibitem{Zhang2018}
M.~Zhang, Y.~Zhang, L.~Zhang, C.~Liu, and S.~Khurshid, ``{DeepRoad}:
  {GAN}-based metamorphic testing and input validation framework for autonomous
  driving systems,'' in \emph{Proceedings of the 33rd ACM/IEEE International
  Conference on Automated Software Engineering (ASE)}, 2018, pp. 132--142.

\bibitem{spieker2019adaptive}
H.~Spieker and A.~Gotlieb, ``Adaptive metamorphic testing with contextual
  bandits,'' \emph{Journal of Systems and Software}, vol. 165, 2020.

\bibitem{tian2018deeptest}
Y.~Tian, K.~Pei, S.~Jana, and B.~Ray, ``{DeepTest}: Automated testing of
  deep-neural-network-driven autonomous cars,'' in \emph{Proceedings of the
  40th Int. Conf. on Soft. Eng. (ICSE)}, 2018, pp. 303--314.

\bibitem{Papadakis2018a}
M.~Papadakis, M.~Kintis, J.~Zhang, Y.~Jia, Y.~L. Traon, and M.~Harman,
  ``Mutation {{Testing Advances}}: {{An Analysis}} and {{Survey}},'' in
  \emph{Advances in {{Computers}}}, 2018.

\bibitem{Jahangirova2020}
G.~Jahangirova and P.~Tonella, ``An {{Empirical Evaluation}} of {{Mutation
  Operators}} for {{Deep Learning Systems}},'' in \emph{{{IEEE International
  Conference}} on {{Software Testing}}, {{Verification}} and {{Validation}}
  ({{ICST}})}, 2020.

\bibitem{Zhu2018DeepMutationMT}
M.~Zhu, F.~Zhang, J.~Sun, M.~Xue, B.~Li, F.~Juefei-Xu, C.~Xie, L.~Li, Y.~Liu,
  J.~Zhao, and Y.~Wang, ``{DeepMutation}: Mutation testing of deep learning
  systems,'' in \emph{IEEE 29th International Symposium on Software Reliability
  Engineering (ISSRE)}, 2018, pp. 100--111.

\bibitem{Wang2019c}
J.~Wang, G.~Dong, J.~Sun, X.~Wang, and P.~Zhang, ``Adversarial sample detection
  for deep neural network through model mutation testing,'' in \emph{Proc. of
  the 41st {{Int. Conf.}} on {{Soft. Eng.}}}, 2019, pp. 1245--1256.

\bibitem{kuhn2009combinatorial}
R.~Kuhn, R.~Kacker, Y.~Lei, and J.~Hunter, ``Combinatorial software testing,''
  \emph{Computer}, vol.~42, no.~8, pp. 94--96, 2009.

\bibitem{kuhn2004software}
D.~R. Kuhn, D.~R. Wallace, and A.~M. Gallo, ``Software fault interactions and
  implications for software testing,'' \emph{IEEE Transactions on Software
  Engineering}, vol.~30, no.~6, pp. 418--421, 2004.

\bibitem{ma2019deepct}
L.~Ma, F.~Juefei-Xu, M.~Xue, B.~Li, L.~Li, Y.~Liu, and J.~Zhao, ``{DeepCT}:
  Tomographic combinatorial testing for deep learning systems,'' in \emph{2019
  IEEE 26th International Conference on Software Analysis, Evolution and
  Reengineering (SANER)}.\hskip 1em plus 0.5em minus 0.4em\relax IEEE, 2019,
  pp. 614--618.

\bibitem{chen2019variable}
Y.~Chen, Z.~Wang, D.~Wang, C.~Fang, and Z.~Chen, ``Variable strength
  combinatorial testing for deep neural networks,'' in \emph{2019 IEEE
  International Conference on Software Testing, Verification and Validation
  Workshops (ICSTW)}.\hskip 1em plus 0.5em minus 0.4em\relax IEEE, 2019, pp.
  281--284.

\bibitem{herbold2020smoke}
S.~Herbold and T.~Haar, ``Smoke testing for machine learning: Simple tests to
  discover severe defects,'' \emph{arXiv preprint arXiv:2009.01521}, 2020.

\bibitem{szegedy2013intriguing}
\BIBentryALTinterwordspacing
C.~Szegedy, W.~Zaremba, I.~Sutskever, J.~Bruna, D.~Erhan, I.~Goodfellow, and
  R.~Fergus, ``Intriguing properties of neural networks,'' in \emph{Int. Conf.
  on Learning Representations}, 2014. [Online]. Available:
  \url{http://arxiv.org/abs/1312.6199}
\BIBentrySTDinterwordspacing

\bibitem{nguyen2015deep}
A.~Nguyen, J.~Yosinski, and J.~Clune, ``Deep neural networks are easily fooled:
  High confidence predictions for unrecognizable images,'' in \emph{Proc. of
  the IEEE conf. on comp. vision and pattern recogn.}, 2015, pp. 427--436.

\bibitem{papernot2016technical}
N.~Papernot, F.~Faghri, N.~Carlini, I.~Goodfellow, R.~Feinman, A.~Kurakin,
  C.~Xie, Y.~Sharma, T.~Brown, A.~Roy \emph{et~al.}, ``Technical report on the
  cleverhans v2. 1.0 adversarial examples library,'' \emph{arXiv preprint
  arXiv:1610.00768}, 2016.

\bibitem{papernot2016limitations}
N.~Papernot, P.~McDaniel, S.~Jha, M.~Fredrikson, Z.~B. Celik, and A.~Swami,
  ``The limitations of deep learning in adversarial settings,'' in \emph{2016
  IEEE European Symposium on Security and Privacy (EuroS\&P)}, 2016, pp.
  372--387.

\end{thebibliography}
\end{document}